\documentclass[twocolumn]{book}
\usepackage[dvips]{graphicx,color}
\usepackage{makeidx,universe}

\makeauthorindex

\BookTitle{Proceedings of the XXIX PHYSICS IN COLLISION}

\CopyRight{\copyright 2009 by Universal Academy Press, Inc.}

\begin{document} 

\pagenumbering{arabic}

\chapter{%
{\LARGE \sf
Search for the Standard Model Higgs Boson at CMS} \\
{\normalsize \bf 
Tommaso Dorigo$^{1}$ (for the CMS collaboration)} \\
{\small \it \vspace{-.5\baselineskip}
(1) INFN, Italy\\
}
}


\AuthorContents{T.\ Dorigo}

\AuthorIndex{Dorigo}{T.}

  \baselineskip=10pt 
  \parindent=10pt    

\section*{Abstract} 
\noindent The prospects for the search of the Standard Model Higgs boson with the CMS experiment at the LHC are presented. The analyses rely on a full simulation of the detector response and emphasis is put on explicit strategies for the measurement of experimental and background systematics from data. The discovery reach is presented as a function of the Higgs boson mass. A new complete strategy is presented for the early searches and for the control of systematics at very low luminosities 
of $O(1 fb^{-1})$.
\section{Introduction} 

\noindent
The Standard Model (SM) requires the existence of a scalar Higgs boson to break electroweak symmetry and provide mass terms to gauge bosons and fermion fields. Indirect constraints from radiative corrections to electroweak observables indicate for the Higgs boson mass the bound $M_H<157\,GeV$~\cite{lepewwg}, at 95\% of Confidence Level (CL). Direct constraints from experimental searches at LEP II and Tevatron have determined that $M_H>114.4\,GeV$, with the exclusion of the range $160\,GeV<M_H<170\,GeV$, at 95\% CL~\cite{tev}. 

\begin{figure}[ht]
   \begin{center}
     \includegraphics[height=15pc,angle=-90]{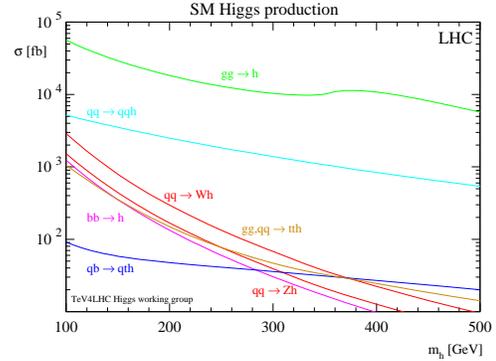}
    \label{fig:Prod}
   \end{center}
  \vspace{-1pc}
  \caption{Higgs production cross section in proton-proton collisions at $14\,TeV$ centre-of-mass energy, in femtobarns, as a function of $M_H$~\cite{tevlhc}. The green curve shows the cross section of the gluon-fusion mechanism, the cyan curve the cross section of Higgs production by vector boson fusion. }
\end{figure}

The SM Higgs boson can be produced in proton-proton collisions at the Large Hadron Collider (LHC) by several different mechanisms (Fig.~\ref{fig:Prod}). The production by gluon-gluon fusion is the most frequent, with cross sections up to several tens of picobarns; smaller is the production by vector boson fusion (VBF), which however provides a striking signature of two forward quark jets. The two main production mechanisms have been investigated to assess the chances of an early detection of the Higgs boson with the Compact Muon Solenoid (CMS) detector~\cite{cms}.

\vspace{3mm}
Searches in three of the main signatures of Higgs production and decay are summarized below; these target the whole favourable range of $M_H$ mass values, with the first two extending above $200\,GeV$ and the third one covering the low-mass region of $M_H<135\,GeV$: 
                       \vspace{-.5\baselineskip}
\begin{itemize}
\item $gg\rightarrow H\rightarrow WW^{(*)}$, with the decay of both $W$ bosons to $e\nu$ or $\mu\nu$ pairs;
                       \vspace{-.5\baselineskip}
\item $gg\rightarrow H\rightarrow ZZ^{(*)}$, with the decay of both $Z$ bosons to $e^+e^-$ or $\mu^+\mu^-$ pairs;
                       \vspace{-.5\baselineskip}
\item $qq\rightarrow qqH$, with a decay $H\rightarrow\tau\tau$ accompanied by two forward hadronic jets.
\end{itemize}
                       \vspace{-.5\baselineskip}

\section{The $H\rightarrow WW$ search}

\noindent
The search for the $WW^{(*)}$ decay mode at CMS~\cite{WW} employs events containing exactly two opposite-charge leptons ($e$ or $\mu$) with transverse momenta $p_T>10 \,GeV$  and pseudorapidity $|\eta|<2.5$, with at least one of them having $p_T>20\,GeV$. The following additional pre-selection cuts are then applied: a jet veto ($N_{jet}^{E_T>15\,GeV}=0$), large missing energy ($E_T^{miss}>30\,GeV$), and a dilepton mass above resonances ($m_{ll}>12\,GeV$). 

\begin{figure}[]
   \begin{center}
     \includegraphics[height=14pc]{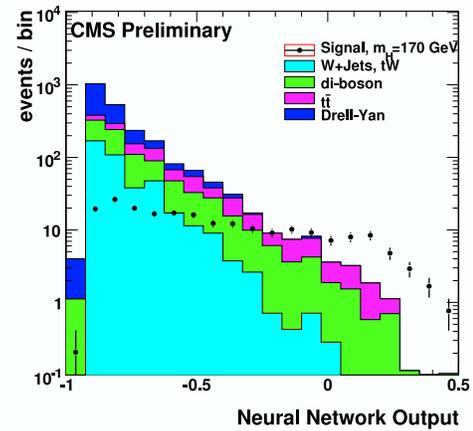}
      \caption{Output of the Neural Network for $H \to WW$ signal (black points) and backgrounds (histograms) for a Higgs boson mass of $170 \,GeV$. A further optimized cut on the NN output is used to select the final candidates.}
    \label{fig:WW_NN}
   \end{center}


\end{figure}

Two separate search strategies are studied: a cut-based analysis and a Neural-Network-based analysis (NN). In both cases the selection is optimized using the azimuthal angle between the leptons, an upper cut on the dilepton mass, and requirements on lepton momenta and missing energy. The NN analysis uses additional kinematic variables to separate the signal from the main backgrounds (Fig.~\ref{fig:WW_NN}).

The analyses include complete techniques to determine background rates with control samples of data. The top-pair background can be sized up with events containing two additional jets, while the SM production of $WW$ pairs can be normalized using data with $m_{ll}>115 \,GeV$. 

The modified frequentist $CL_s$ method \cite{tevlhc} is used to convert the number of expected signal and background events into a significance of the observable signal, as a function of the Higgs mass. A first evidence for the Higgs boson is likely achievable in $1 fb^{-1}$ with the $WW$ final state alone in the region of best sensitivity, $155<M_H<180$ GeV. 

\begin{figure}[]
   \begin{center}
     \includegraphics[height=13pc]{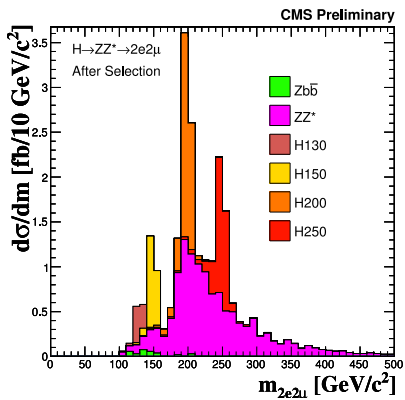}
      \caption{Effective cross section as a function of the four-lepton invariant mass after the signal selection, in the mixed $ee\mu\mu$ channel of the $H\to ZZ$ search. Four different signal distributions (for masses of $130,\,150,\,200$ and $250\,GeV$) are compared to residual backgrounds. The latter are essentially due to SM $ZZ$ production (in purple) and $Zbb$ production (in green).}
    \label{fig:ZZ_xsection}
	\end{center}
\end{figure}

\section{The $H\rightarrow ZZ$ search}

\noindent
In the $H\rightarrow ZZ^{(*)}$ search~\cite{ZZ} events are selected to contain four charged leptons ($e^+e^-e^+e^-$, $e^+e^-\mu^+\mu^-$, or $\mu^+\mu^-\mu^+\mu^-$), with pair masses $m_{ll}>12 \,GeV$. To remove the $Zb\bar b$ and $t\bar t$ backgrounds further, CMS uses the combined isolation of the two least-isolated leptons, and the significance of their impact parameter with respect to the primary vertex. The reconstructed mass of the dilepton pairs is requested to lay in the windows $[50-100]$ and $[20-100] \,GeV$. 

\begin{figure}[h!]
   \begin{center}
     \includegraphics[height=13pc]{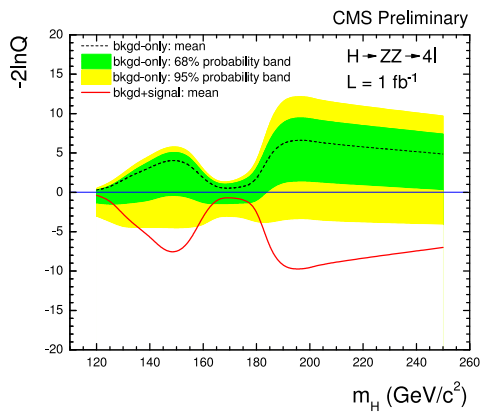}
      \caption{Values of the log-likelihood ratio discriminant resulting from the combination of the four-lepton final states studied in the search. The dashed black curve shows the expected distribution of the discriminant in the absence of signal as a function of Higgs mass, while the red curve shows the observable value of the ratio in the presence of the Higgs boson.}
    \label{fig:ZZ_likelihood}
   \end{center}
\end{figure}

\noindent
After the selection, backgrounds are almost exclusively due to SM $ZZ^{(*)}$ production. The four-body mass provides further discrimination, as shown in Fig.~\ref{fig:ZZ_xsection} Mass window cuts allow to obtain a signal-to-noise ratio larger than 1.0 throughout the $120-250 \,GeV$ Higgs mass region considered in the search. 

The $CL_s$ method allows to estimate the significance of the extractable signal  with $1 fb^{-1}$ of $14 TeV$ collision (Fig.~\ref{fig:ZZ_likelihood}). A sensitivity at the $2\sigma$ level can be obtained for favourable mass values. 

\section{The $qqH\rightarrow qq\tau\tau$ search}

\noindent
A study of the observability of the VBF signature $qqH\rightarrow qq\tau\tau$, for Higgs masses between $115$ and $145 \,GeV$, has been performed~\cite{taus} using events containing one leptonic $\tau$-decay candidate ($\tau\rightarrow e\nu\nu$ or $\tau\rightarrow \mu\nu\nu$), collected by a low-$p_T$ electron or muon trigger. A second $\tau$-lepton candidate is required to produce a narrow $E_T>30 \,GeV$ jet containing one track with $p_T>6\,GeV$ within its core. 

The two forward jets characteristic of VBF processes  are used to reduce backgrounds, mainly coming from QCD multijet production and $Z\rightarrow \tau\tau$ decays. The mass of two forward jets with $E_T>30\,GeV$ has to exceed $400\,GeV$, and they must be separated in pseudorapidity by more than $2.5$ units. 

Backgrounds amount to $31.8$ events, with $0.6-1.6$ expected from the Higgs signal, depending on $M_H$. 
The sensitivity of this search channel is found to be insufficient to provide an independent evidence of the SM $H$ boson in early LHC data.




\section{Combination of $H\rightarrow ZZ$, $H\rightarrow WW$ searches}

\noindent
An additional study was carried out for a combination of the $WW$ and $ZZ$ channels to determine the range of Higgs boson masses that CMS is likely to exclude at 95\% C.L. in the absence of a signal, using the results of~\cite{WW} and~\cite{ZZ}. The combination was performed with both the $CL_s$ and a Bayesian method; in general the two methods were found to agree within 10\%, which is also a measure of the typical variation in their difference. 

\begin{figure}[h!]
  \begin{center}
    \includegraphics[height=15pc]{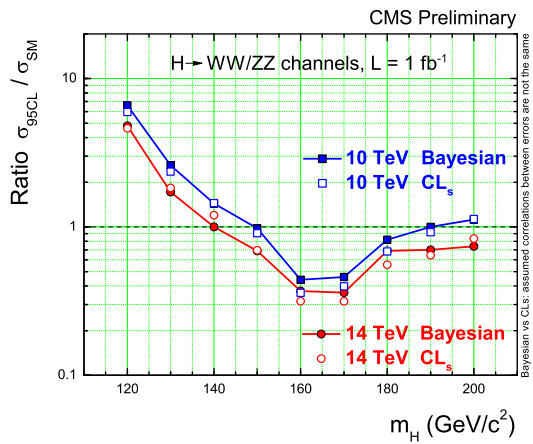} 
    \label{fig:comb}
  \end{center}
  \vspace{-1pc}
  \caption{Predicted limits on $H$ cross section in units of $\sigma_{SM}$ as a function of $M_H$  for $1 fb^{-1}$ at $10$ and $14\,TeV$, obtained by combining $H \to WW$ and $H \to ZZ$ search results.}
\end{figure}

Fig.~\ref{fig:comb} shows the limit which can be obtained with luminosity of $1 fb^{-1}$ at $14\,TeV$ together with the result of considering a modified scenario, in which $1 fb^{-1}$ of collisions is produced at the reduced energy of $10\,TeV$.

\section{Acknowledgements}

\noindent
The author wishes to thank Eleni Petrakou for her editorial help in the preparation of this paper.



%
%

\end{document}